\documentstyle[aps]{revtex}
\tighten
\begin{document}


\preprint{hep-th/0104185}
\title{Integro-differential equation for brane-world cosmological
perturbations} 
\author{Shinji Mukohyama}
\address{
Department of Physics and Astronomy, University of Victoria\\ 
Victoria, BC, Canada V8W 3P6
}

\maketitle


\begin{abstract} 

Cosmological perturbations in the brane-world cosmology with a
positive tension brane in the AdS background bulk geometry is analyzed 
by using the doubly gauge-invariant formalism. We derive four
independent equations for scalar perturbations in the plane symmetric
($K=0$) background. Three of these equations are differential
equations written in terms of gauge invariant variables on the brane
only, and another of them is an integro-differential equation whose
kernel is constructed formally from the ${\bf Z}_2$-symmetric retarded
Green's function of the bulk gravitational waves. We compare these
four equations with the corresponding equations in the standard
cosmology. As a by-product, we also obtain a set of equations which
may be useful in numerical calculations. 

\end{abstract}



\section{Introduction}


The idea that our four-dimensional world may be a timelike surface, or 
a world-volume of a $3$-brane, in a higher dimensional spacetime has
been attracting a great deal of physical interests. As shown by
Randall and Sundrum~\cite{RS2}, the $4$-dimensional Newton's law of
gravity can be reproduced on a $4$-dimensional timelike hypersurface
with positive tension in a $5$-dimensional AdS background despite the
existence of the infinite fifth dimension.


Moreover, in refs.~\cite{CGS,FTW,BDEL,Mukohyama2000a,Kraus,Ida} it was 
shown that the standard cosmology can be realized in the
scenario in low energy as far as a spatially homogeneous and isotropic
brane is concerned. Hence, it seems interesting to look for observable 
consequences of the cosmology based on this scenario. For this
purpose, cosmic microwave background (CMB) anisotropy seems a powerful
tool. In this respect, many authors investigated cosmological
perturbations in the brane-world 
scenario~\cite{Mukohyama2000b,Mukohyama2000c,KIS,Kodama,Maartens,Langlois,BDBL,Koyama-Soda,LMW,LMSW}.


In this paper, we investigate classical evolution of brane-world
cosmological perturbations. It was conjectured in
ref.~\cite{Mukohyama2000b} that the evolution of cosmological 
perturbations becomes non-local in the brane-world in the sense that
it should be described by some integro-differential equations. The
non-locality is due to gravitational waves propagating in the bulk
$5$-dimensional spacetime. Thus, the main purpose of this paper is to
give a formal derivation of an integro-differential equation for
brane-world cosmological perturbations. Actually, we shall derive four
independent equations for scalar perturbations on the brane, one of
which is the integro-differential equation, by using the doubly
gauge-invariant formalism developed by the 
author~\cite{Mukohyama2000b,Mukohyama2000c}.


In section~\ref{sec:background} we review the background cosmological
solutions in the brane-world scenario and describe the background we
use in this paper. In section~\ref{sec:formalism}, specializing to
scalar perturbations around a plane-symmetric ($K=0$) background, we
review the doubly gauge invariant formalism of brane-world
cosmological perturbations. Section~\ref{sec:eq-on-brane} is the main
part of this paper. In the section we derive the four equations for
perturbatious on the brane including the integro-differential
equation. In section~\ref{sec:comparison} we compare these four
equations in the brane-world cosmology with the corresponding
equations in the standard cosmology. Finally,
section~\ref{sec:summary} is devoted to a summary of this paper.


\section{Background geometry}
	\label{sec:background}

As mentioned in the introduction, cosmological solutions in the 
Randall-Sundrum brane world scenario were recently 
found~\cite{BDEL,Mukohyama2000a,Kraus,Ida}. In these
solutions, the standard cosmology is restored in low energy, if a 
parameter in the solutions is small enough. If the parameter is not
small enough, it affects cosmological evolution of our universe as
dark radiation~\cite{Mukohyama2000a}. Hence, the parameter should be
very small in order that the brane-world scenario should be consistent
with nucleosynthesis~\cite{BDEL}. On the other hand, in
ref.~\cite{MSM}, it was shown that $5$-dimensional geometry of all
these cosmological solutions is the Schwarzschild-AdS (S-AdS)
spacetime~\cite{Birmingham} and that the parameter is equivalent to 
the mass of the black hole. Therefore, the $5$-dimensional bulk 
geometry should be the S-AdS spacetime with a small mass, which is
close to the pure AdS spacetime. Moreover, black holes with small mass
will evaporate in a short time scale~\cite{Hawking}. Thus, it seems a
good approximation to consider the pure AdS spacetime as a
$5$-dimensional bulk geometry for the brane-world cosmology.

Now let us review the above by using several equations and describe
the background geometry which we shall consider throughout this
paper. First, let us consider a five-dimensional bulk metric of the
form
%
\begin{equation}
 ds_5^2 = -N^2(\tau,w)d\tau^2 + r^2(\tau,w)d\Sigma^2_K + dw^2,
	\label{eqn:Gaussian-metric}
\end{equation}
where $d\Sigma^2_K$ is a metric of a unit three-dimensional sphere,
plane or hyperboloid for $K=+1,0$ or $-1$, respectively. This metric
represents a general five-dimensional spacetime with the symmetry of
four-dimensional Friedmann universe. Actually, the induced metric on
the constant-$w$ hypersurface ($w=w_0$) is of the four-dimensional
Friedmann metric, provided that the new time coordinate $t$ and the
scale factor $a(t)$ are introduced by $dt=N(\tau,w_0)d\tau$ and
$a(t)=r(\tau,w_0)$.  Hence, the metric (\ref{eqn:Gaussian-metric})
represents the bulk geometry for the brane-world cosmology in the
Gaussian normal coordinate system and $w$ represents the extra
dimension if we impose the $Z_2$-symmetry, or the invariance under the 
parity transformation $w-w_0\to w_0-w$. It is possible to solve the 
five-dimensional Einstein equations with the negative cosmological
constant by using this form of the metric. The bulk solution and the
corresponding cosmological evolution equation induced on the brane
were obtained in refs.~\cite{BDEL,Mukohyama2000a}.

Instead, we may consider a different coordinate system in which the
metric is of the following form~\footnote{
For latter convenience we included $l^2$ in front of $dT^2$. So, the
time variable $T$ in the present paper corresponds to $T/l$ in
ref.~\cite{MSM}.}. 
%
\begin{equation}
 ds_5^2 = -F(T,r) l^2dT^2 + G(T,r)dr^2 + r^2 d\Sigma^2_K. 
	\label{eqn:metric-S-AdS}
\end{equation}
In this coordinate system, the five-dimensional Einstein equations
with the negative cosmological constant can be solved more easily than 
in the Gaussian normal coordinate. The solution is the S-AdS
spacetime~\cite{Birmingham}:
%
\begin{equation}
 F = G^{-1} = K + \frac{r^2}{l^2} - \frac{\mu}{r^2},
	\label{eqn:S-AdS}
\end{equation}
where $l$ is the length scale of the five-dimensional negative
cosmological constant and $\mu$ is a constant. This statement may be
considered as a generalized Birkoff's theorem~\footnote{
Of course, if we consider matter fields~\cite{GW,DFGK} or quantum
corrections~\cite{Mukohyama2001a,Mukohyama2001b} in the bulk then the
generalized Birkoff's theorem does not hold. In these more general
case, the bulk geometry becomes completely dynamical if the induced
geometry on the brane is dynamical (eg. expanding). Thus, inclusion of
bulk matter fields or quantum corrections is beyond the scope of this
paper.
}. Thus, the bulk geometry is static, regardless of the dynamical
motion of the brane, insofar as the world-volume of the brane has the 
symmetry of the three-dimensional sphere, plane or hyperboloid,
respectively. The $Z_2$-symmetry is incorporated by considering two
copies of the same region in the S-AdS spacetime with a timelike
boundary and gluing these along the boundary. The corresponding
cosmological evolution equation induced on the brane was obtained in
refs.~\cite{Kraus,Ida} and is the same as that obtained by using the
Gaussian normal coordinate system.

It is, of course, possible to obtain explicit coordinate
transformation between the Gaussian normal coordinate system and the
spherically, plane or hyperbolically symmetric coordinate
system. Actually, in ref.~\cite{MSM} the coordinate transformation was 
obtained and the global structure of the solutions was investigated by
using it.

As pointed out in ref.~\cite{Mukohyama2000a} and explained in the
first paragraph of this section, consideration of the dark radiation
and the Hawking radiation makes us to think that the pure AdS
spacetime, which corresponds to $\mu=0$, should be good enough for the
bulk geometry. The generalized Birkoff's theorem guarantees that the
AdS bulk spacetime is perfectly consistent with any motion of the
brane, provided that the induced geometry on the brane has the
symmetry of the three-dimensional sphere, plane or hyperboloid. 
Hence, throughout this paper, we shall consider the pure AdS spacetime
as the bulk geometry. Furthermore, for simplicity we shall consider
the spatially flat ($K=0$) brane only. In this case, it is convenient
to introduce a new spatial coordinate $X$ by $X=l/r$ so that the
metric (\ref{eqn:metric-S-AdS}) with $\mu=K=0$ becomes manifestly
conformally flat: 
%
\begin{eqnarray}
 ds_5^2 = g^{(0)}_{MN}dx^Mdx^N = \left(\frac{l}{X}\right)^2\left[
	-dT^2 + dX^2 + \sum_{i=1}^3(dx^i)^2\right],
		\label{eqn:metric-ads}
\end{eqnarray}
where $M,N=0,1,2,3,4$.

In this bulk spacetime, let us consider a world volume of a $3$-brane
or, a timelike hypersurface, $\Sigma^{(0)}$ given by the imbedding
relation $x^M=Z^{(0)M}(y)$, where $y$ represents $\{y^{\mu}\}$
($\mu=0,1,2,3$) and 
%
\begin{eqnarray}
 Z^{(0)T} & = & \bar{T}(t), \nonumber\\
 Z^{(0)X} & = & \bar{X}(t), \nonumber\\
 Z^{(0)i} & = & y^i\quad (i=1,2,3). 
	\label{eqn:imbedding-ads}
\end{eqnarray}
Hereafter, $t=y^0$. The induced metric $q^{(0)}_{\mu\nu}$ on
$\Sigma^{(0)}$ defined by 
%
\begin{eqnarray}
 q^{(0)}_{\mu\nu}(y) & = & 
	\left.g^{(0)}_{MN} 
	e^{(0)M}_{\mu}e^{(0)N}_{\nu}\right|_{x=Z^{(0)}(y)},\nonumber\\ 
 e^{(0)M}_{\mu} & = & \frac{\partial Z^{(0)M}}{\partial y^{\mu}},
\end{eqnarray}
can be calculated to be
%
\begin{equation}
 q^{(0)}_{\mu\nu}dy^{\mu}dy^{\nu} 
	= -dt^2 + a^2(t)\sum_{i=1}^3(dy^i)^2,
\end{equation}
where the time coordinate $t=y^0$ has been normalized so that 
$(l\dot{\bar{T}}/\bar{X})^2-(l\dot{\bar{X}}/\bar{X})^2=1$ 
and the function $a(t)$ is defined by $a(t) = l/\bar{X}(t)$. This is
actually the flat Friedmann-Robertson-Walker metric. The extrinsic 
curvature of $\Sigma^{(0)}$ defined by 
%
\begin{equation}
 K^{(0)}_{\mu\nu}(y) = \left.\frac{1}{2}e^{(0)M}_{\mu}e^{(0)N}_{\nu}
	{\cal L}_{n^{(0)}}g^{(0)}_{MN}\right|_{x=Z^{(0)}(y)}, 
\end{equation}
where $n^{(0)M}$ is the unit normal to $\Sigma^{(0)}$, is calculated
to be 
%
\begin{eqnarray}
 K^{(0)}_{\mu\nu}dy^{\mu}dy^{\nu} & = & {\cal K}(t)dt^2
        + a^2(t)\bar{\cal K}(t)\sum_{i=1}^3(dy^i)^2, \nonumber\\
 {\cal K}(t) & = & 
	\frac{l^2H(t)^2+l^2\dot{H}(t)+1}{l\sqrt{1+l^2H(t)^2}}, 
	\nonumber\\
 \bar{\cal K}(t) & = & -\frac{\sqrt{1+l^2H(t)^2}}{l}.
	\label{eqn:unperturbed-K}
\end{eqnarray}
Here, $H$ is defined by $H=\dot{a}/a$ and the direction of the unit 
normal vector $n^M$ has been chosen so that 
$n^{(0)M}\partial_Mr|_{x=Z^{(0)}(y)}<0$.

The $Z_2$-symmetry can be incorporated by considering two copies of
one of the two regions (i) $X>\bar{X}(\bar{T}^{-1}(T))$ or (ii)
$X<\bar{X}(\bar{T}^{-1}(T))$, and gluing the two copies along 
$\Sigma^{(0)}$. Hence, Israel's junction condition~\cite{Israel} gives
the evolution equation of $a(t)$ as  
%
\begin{equation}
 \pm\frac{\sqrt{1+l^2H(t)^2}}{l} =
	\frac{\kappa_5^2}{6}(\rho(t)+\lambda)
	\label{eqn:Friedmann-eq-pre}
\end{equation}
and the conservation equation
%
\begin{equation}
 \dot{\rho}(t) + 3H(t)(\rho(t)+p(t)) = 0,
	\label{eqn:background-conservation}
\end{equation}
where $\kappa_5$ is the $5$-dimensional gravitational constant and we
have assumed that the surface energy-momentum tensor, or the Lanczos
tensor, $S^{(0)}_{\mu\nu}$ of the brane is of the form 
%
\begin{equation}
 S^{(0)}_{\mu\nu} = -\lambda q_{\mu\nu} + T_{\mu\nu},
\end{equation}
where $\lambda$ is the vacuum energy density on the brane, or the 
tension of the brane, and 
%
\begin{equation}
 T^{\mu}_{\nu} = \left(\begin{array}{cccc}
        -\rho(t) & 0 & 0 & 0 \\
        0 & p(t) & 0 & 0 \\
        0 & 0 & p(t) & 0 \\
        0 & 0 & 0 & p(t) 
        \end{array}\right). 
	\label{eqn:Tmunu}
\end{equation}
The sign in (\ref{eqn:Friedmann-eq-pre}) is plus or minus for the
choice (i) or (ii), respectively. 
In order for the vacuum $\rho=p=0$ to give the nearly static brane 
$l^2H^2\ll 1$, it is necessary and sufficient that
$\kappa_5^2\lambda\simeq\pm 6/l$, where the plus and the minus 
signs are for the choice (i) and (ii), respectively. Hence, the
tension $\lambda$ of the brane should be positive (or negative) for 
the choice (i) (or (ii), respectively). With this sign of $\lambda$, 
the evolution equation (\ref{eqn:Friedmann-eq-pre}) can be rewritten 
as 
%
\begin{equation}
 H^2 = \frac{8\pi G_N}{3}\left( 1+ 
        \frac{\rho}{2\lambda}\right)\rho
        +\frac{\Lambda_4}{3}, 
	\label{eqn:Friedmann-eq}
\end{equation}
where $G_N=\kappa_5^4 \lambda /48\pi$ and
$\Lambda_4=\kappa_5^4\lambda^2/12-3l^{-2}$. 
Hence, for the choice (i), we obtain the standard Friedmann equation
in low energy (namely when $|\rho|\ll\lambda$) and the evolution
equation (\ref{eqn:Friedmann-eq}) can be considered as a generalized
Friedmann equation~\cite{CGS,FTW,BDEL,Mukohyama2000a,Kraus,Ida}. 
On the other hand, if we choose (ii) then the effective
four-dimensional Newton's constant $G_N$ is negative and, thus, the
cosmology in low energy becomes very different from the standard 
cosmology~\cite{SMS}~\footnote{
If we consider matter fields or quantum effects in the bulk then this
conclusion might be significantly changed since the near brane
geometry in the bulk can be very different from AdS spacetime. For
analysis of weak gravity in models with bulk matter fields, see
ref.~\cite{TM}.}. Therefore, in the following we shall consider the
choice (i). Namely, we shall consider two copies of the region
$X>\bar{X}(\bar{T}^{-1}(T))$ in the pure AdS spacetime and glue them
along the trajectory $(T,X)=(\bar{T}(t),\bar{X}(t))$ of the brane with
positive $\lambda$.


\section{Doubly gauge-invariant formalism}
	\label{sec:formalism}

In this section, specializing to scalar perturbations~\footnote{
This treatment is, of course, justified since vector and tensor
perturbations are decoupled from the scalar perturbations in the
linear order. See ref.~\cite{Mukohyama2000a} for the
categorization.} around a plane-symmetric ($K=0$) background, we
review the gauge-invariant formalism of gravitational perturbations in
the bulk and the doubly gauge-invariant formulation of the perturbed
junction condition developed in
refs.~\cite{Mukohyama2000b,Mukohyama2000c}. In the next section we 
shall apply the formalism to the brane-world cosmological background
described in the previous section.

\subsection{Perturbed junction condition}

First, let us consider a general five-dimensional metric $g_{MN}$,
general functions $\{Z^M(y)\}$ determining the imbedding of a general 
world volume $\Sigma$ of the brane, where $y$ denotes $4$ dimensional 
coordinates $\{y^{\mu}\}$ on $\Sigma$, and a general surface
energy-momentum tensor $S_{\mu\nu}$ on $\Sigma$. We then expand them
around a general background specified by $g^{(0)}_{MN}$,
$\{Z^{(0)M}(y)\}$ and $S^{(0)}_{\mu\nu}$. Namely, we  consider the
metric 
%
\begin{equation}
 ds_5^2 = g_{MN}dx^Mdx^N = (g_{MN}^{(0)} + \delta g_{MN})dx^Mdx^N,
\end{equation}
the imbedding relation 
%
\begin{equation}
 x^M = Z^M(y) = Z^{(0)M}(y) + \delta Z^M(y),
\end{equation}
and the surface energy momentum tensor
%
\begin{equation}
 S_{\mu\nu} = S^{(0)}_{\mu\nu}+\delta S_{\mu\nu}. 
\end{equation}
We can calculate the perturbed induced metric and the perturbed
extrinsic curvature of $\Sigma$ up to the linear order and obtain 
the following expressions. 
%
\begin{eqnarray}
 q_{\mu\nu} & = & q^{(0)}_{\mu\nu}+\delta q_{\mu\nu}, \nonumber\\
 K_{\mu\nu} & = & K^{(0)}_{\mu\nu}+\delta K_{\mu\nu}, 
	\label{eqn:perturb-q-K}
\end{eqnarray}
where
%
\begin{eqnarray}
 \delta q_{\mu\nu} & = & e^{(0)M}_{\mu}e^{(0)N}_{\nu}
        (\delta g_{MN} + {\cal L}_{\delta Z}g^{(0)}_{MN}),
        \nonumber\\
 \delta K_{\mu\nu} & = &
        \frac{1}{2}n^{(0)M}n^{(0)N}
        (\delta g_{MN}+2\delta Z_{M;N}) K^{(0)}_{\mu\nu}        
        \nonumber\\
 & &    - \frac{1}{2}n^{(0)L}e^{(0)M}_{\mu}e^{(0)N}_{\nu}
        \left[ 2\delta\Gamma_{LMN}
        + \delta Z_{L;MN} + \delta Z_{L;NM}
        + (R^{(0)}_{L'MLN}+R^{(0)}_{L'NLM})\delta Z^{L'}\right].
	\label{eqn:delta-q-K}
\end{eqnarray}
Here, ${\cal L}$ denotes the Lie derivative defined in the 
five-dimensional spacetime, the semicolon denotes the covariant 
derivative compatible with the background metric $g^{(0)}_{MN}$, 
$R^{(0)}_{L'MLN}$ is the Riemann tensor of $g^{(0)}_{MN}$ and
$\delta\Gamma_{LMN}=
(1/2)(\delta g_{LM;N}+\delta g_{LN;M}-\delta g_{MN;L})$. 
The perturbations $\delta q_{\mu\nu}$ and $\delta K_{\mu\nu}$ 
may be evaluated on $\Sigma^{(0)}$.

For these perturbations we have the following two kinds of gauge
transformations~\cite{Mukohyama2000c}. We have the {\it $5$-gauge
transformation} (infinitesimal coordinate transformation in the
$5$-dimensional bulk spacetime), 
$x^M \to {x'}^M=x^M+\bar{\xi}^M(x)$, under which $\delta g_{MN}$ and
$\delta Z^M$ transform as 
%
\begin{eqnarray}
 \delta g_{MN} & \to &
        \delta g_{MN} -\bar{\xi}_{M;N} -\bar{\xi}_{N;M},
        \nonumber\\
 \delta Z^M & \to & \delta Z^M + \bar{\xi}^M,
        \label{eqn:5-gauge-tr}
\end{eqnarray}
but $\delta q_{\mu\nu}$, $\delta K_{\mu\nu}$ and $\delta S_{\mu\nu}$
are invariant. We also have the {\it $4$-gauge transformation}
(infinitesimal reparameterization of the $4$-dimensional hypersurface
$\Sigma$), $y^{\mu}\to{y'}^{\mu}=y^{\mu}+\bar{\zeta}^{\mu}(y)$, under
which 
%
\begin{eqnarray}
 \delta q_{\mu\nu} & \to & \delta q_{\mu\nu} 
        - \bar{\cal L}_{\bar{\zeta}}q^{(0)}_{\mu\nu}, \nonumber\\
 \delta K_{\mu\nu} & \to & \delta K_{\mu\nu} 
        - \bar{\cal L}_{\bar{\zeta}}K^{(0)}_{\mu\nu}, \nonumber\\
 \delta S_{\mu\nu} & \to & \delta S_{\mu\nu} 
        - \bar{\cal L}_{\bar{\zeta}}S^{(0)}_{\mu\nu}.
        \label{eqn:4-gauge-tr}
\end{eqnarray}
Here, $\bar{\cal L}$ denotes the Lie derivative defined in the
$4$-dimensional manifold $\Sigma^{(0)}$. Note that the $4$-gauge 
transformation is not a part of the $5$-gauge transformation. 
Actually, $\delta q_{\mu\nu}$ and $\delta K_{\mu\nu}$ are invariant 
under the $5$-gauge transformation but are not invariant under the
$4$-gauge transformation. These two gauge transformation were 
disentangled from each other in ref.~\cite{Mukohyama2000c}.

Next, as the background let us consider the plane-symmetric metric 
%
\begin{equation}
 g^{(0)}_{MN}dx^Mdx^N = 
        \gamma_{ab}dx^adx^b + r^2\sum_{i=1}^3(dx^i)^2,
\end{equation}
and such imbedding functions $Z^{(0)M}(y)$ that $Z^{(0)a}$ depends 
only on $y^0$ and that $Z^{(0)i}=y^i$, where the two-dimensional 
metric $\gamma_{ab}$ and the function $r^2$ are assumed to depend 
only on the two dimensional coordinates $\{x^a\}$. Let us call the 
two-dimensional spacetime spanned by $\{x^a\}$ an {\it orbit space} 
since the unperturbed motion of the brane can be described as an 
orbit in this two-dimensional spacetime. As in the previous section, 
we assume that 
the hypersurface $\Sigma^{(0)}$ is timelike and $y^0$ is normalized so
that it represents the proper time $t$ along the trajectory of the
brane. The background metric (\ref{eqn:metric-ads}) and imbedding
functions (\ref{eqn:imbedding-ads}) described in the previous section
is an example. In the plane symmetric background we can decompose 
perturbations as follows 
%
\begin{eqnarray}
 \delta g_{MN}dx^Mdx^N & = & \int d^3{\bf k}\left[
        h_{ab}Ydx^adx^b 
        + 2h_{(L)a}V_{(L)i}dx^adx^i
        \right.\nonumber\\
 & &    \left.
        + (h_{(LL)}T_{(LL)ij}+h_{(Y)}T_{(Y)ij})dx^idx^j\right],
        \nonumber\\
 \delta Z_Mdx^M & = & \int d^3{\bf k}\left[
        z_aYdx^a
        + z_{(L)}V_{(L)i}dx^i\right],
        \nonumber\\
 \delta S_{\mu\nu}dy^{\mu}dy^{\nu} & = & \int d^3{\bf k}\left[
        t_{00}Ydy^0dy^0
        + 2t_{(L)0}V_{(L)i}dy^0dy^i
        \right.\nonumber\\
 & &    \left.
        + (t_{(LL)}T_{(LL)ij}+t_{(Y)}T_{(Y)ij})dy^idy^j\right],
        \label{eqn:harmonic-expansion}
\end{eqnarray}
where $Y=\exp(-i{\bf k}\cdot{\bf x})$, $V_{(L)i}=\partial_iY$,
$T_{(LL)ij}=2\partial_i\partial_jY+(2{\bf k}^2/3)\delta_{ij}Y$ and 
$T_{(Y)ij}=\delta_{ij}Y$, and all coefficients are supposed to depend
only on the $2$-dimensional coordinates $\{x^a\}$ of the orbit space. 
Here, ${\bf x}$ denotes coordinates $\{x^i\}$ of the 
three-dimensional plane ($i=1,2,3$), and ${\bf k}$ represents the 
momentum $\{ k_i\}$ along the plane. Hereafter, we omit ${\bf k}$ 
in most cases. It is easy to see
how the coefficients $\{h's, z's\}$ transform under the $5$-gauge 
transformation by using (\ref{eqn:5-gauge-tr}) and to construct
{\it $5$-gauge-invariant variables}, those linear combinations which 
are invariant under the $5$-gauge transformation. Therefore, we 
obtain $5$-gauge-invariant variables corresponding to perturbations 
of physical position of the hypersurface $\Sigma$ 
%
\begin{equation}
 \phi_a = z_a + X_a, 
\end{equation}
as well as $5$-gauge-invariant variables representing gravitational
perturbations in the bulk
%
\begin{eqnarray}
 F_{ab} & = & h_{ab}-\nabla_aX_b-\nabla_bX_a,
        \nonumber\\
 F & = & h_{(Y)} -X^a\partial_br^2+\frac{2k^2}{n}h_{(LL)},
\end{eqnarray}
where $X_a=h_{(L)a}-r^2\partial_a(r^{-2}h_{(LL)})$.

It is also easy to obtain the corresponding expansion of 
$\delta q_{\mu\nu}$ and $\delta K_{\mu\nu}$ by using the expression
(\ref{eqn:delta-q-K}). After analyzing $4$-gauge transformations
(\ref{eqn:4-gauge-tr}) of coefficients of the expansion, we can
construct {\it $4$-gauge-invariant variables}, those linear 
combinations that are invariant under the $4$-gauge transformation, 
and express them in terms of $5$-gauge-invariant variables as 
follows. For perturbations of the
induced metric, we have~\footnote{In ref.~\cite{Mukohyama2000c}, there
is a typo in the sign of the second term of $f_{00}$.}
%
\begin{eqnarray}
 f_{00} & = & e^ae^bF_{ab} + 2{\cal K}\phi,
        \nonumber\\
 f & = & F + \phi n^a\partial_a r^2,
	\label{eqn:f00-f-Fab-F}
\end{eqnarray}
where $e^a$, $n^a$ and $\phi$ are abbreviations for $e^{(0)a}_0$, 
$n^{(0)a}$ and $n^{a}\phi_a$, respectively, and ${\cal K}$ denotes 
the $00$-component $K^{(0)}_{00}$ of the unperturbed extrinsic 
curvature as in (\ref{eqn:unperturbed-K}). For perturbations of the 
extrinsic curvature, we have
%
\begin{eqnarray}
 k_{00} & = & -\frac{1}{2}n^ae^be^c
        (2\nabla_cF_{ab}-\nabla_aF_{bc})
        +\frac{1}{2}(n^an^b+2e^ae^b)F_{ab}{\cal K}
        - \ddot{\phi}
        +\frac{1}{2}(R^{(\gamma)}+2{\cal K}^2)\phi, 
        \nonumber\\
 k_{(L)0} & = & -\frac{1}{2}n^ae^bF_{ab}
        -r(r^{-1}\phi)^{\cdot},
        \nonumber\\
 k_{(LL)} & = & -\frac{1}{2}\phi,
        \nonumber\\
 k_{(Y)} & = & \frac{1}{4} F_{ab}n^a
        (2e^be^c-n^bn^c)\partial_cr^2
        + \frac{1}{2}r^2n^a\partial_a(r^{-2}F)
	+ \frac{1}{2}\dot{\phi}e^c\partial_cr^2
        + \phi\left( r^2n^bn^c\nabla_b\nabla_c\ln r 
        +\frac{{\bf k}^2}{3}\right),
	\label{eqn:k00-kL0-kLL-kY}
\end{eqnarray}
where $\nabla_a$ is the two-dimensional covariant derivative
compatible with the metric $\gamma_{ab}$ of the orbit space and
$R^{(\gamma)}$ is the Ricci scalar of $\gamma_{ab}$ and dots denote
derivative with respect to the proper time $t$. Note that the right
hand sides of (\ref{eqn:f00-f-Fab-F}) and (\ref{eqn:k00-kL0-kLL-kY})
are evaluated on the unperturbed hypersurface $\Sigma^{(0)}$ and that
$\phi$ can be considered as a function of the proper time $t$.

Now, for completeness, we show the relation among the
$4$-gauge-invariant variables and ($\delta q_{\mu\nu}$, 
$\delta K_{\mu\nu}$) in (\ref{eqn:perturb-q-K}). 
%
\begin{eqnarray}
 f_{00} & = & \sigma_{00}-2\dot{\chi}, \nonumber\\
 f & = & \sigma_{(Y)} + 2\chi r\dot{r}
        + \frac{2k^2}{3}\sigma_{(LL)},
        \nonumber\\
 k_{00} & = & \tilde{\kappa}_{00}
        + \chi\dot{\cal K},	\nonumber\\
 k_{(L)0} & = & \tilde{\kappa}_{(L)0}
        + \frac{1}{2}(\bar{\cal K}+{\cal K})
        \left[\chi 
        - r^2(r^{-2}\sigma_{LL})^{\cdot}\right], 
        \nonumber\\
 k_{(LL)} & = & \tilde{\kappa}_{(LL)}
        \nonumber\\
 k_{(Y)} & = & \tilde{\kappa}_{(Y)}
        + r^2\chi\dot{\bar{\cal K}},
\end{eqnarray}
where $\chi= 
\sigma_{(L)0}-r^2(r^{-2}\sigma_{(LL)})^{\cdot}$ and
%
\begin{eqnarray}
 \delta q_{\mu\nu}dy^{\mu}dy^{\nu} & = & \sum_k\left[
        \sigma_{00}Ydy^0dy^0
        + 2\sigma_{(L)0}V_{(L)i}dy^0dy^i
        \right.\nonumber\\
 & &    \left.
        + (\sigma_{(LL)}T_{(LL)ij}+\sigma_{(Y)}T_{(Y)ij})
	dy^idy^j\right],
        \nonumber\\
 \left[\delta K_{\mu\nu} - \frac{1}{2}
        (K^{(0)\rho}_{\mu}\delta q_{\rho\nu}
        +K^{(0)\rho}_{\nu}\delta q_{\rho\mu})\right]
	dy^{\mu}dy^{\nu} & = & \sum_k\left[
        \tilde{\kappa}_{00}Ydy^0dy^0
        + 2\tilde{\kappa}_{(L)0}V_{(L)i}dy^0dy^i
        \right.\nonumber\\
 & &    \left.
        + (\tilde{\kappa}_{(LL)}T_{(LL)ij}
        +\tilde{\kappa}_{(Y)}T_{(Y)ij})dy^idy^j\right].
\end{eqnarray}

Finally, the doubly gauge-invariant junction condition with the
$Z_2$-symmetry becomes 
%
\begin{eqnarray}
 2k_{00} & = & -\kappa^2\left(
        \frac{2}{3}\tau_{00}+a^{-2}\tau_{(Y)}\right),
        \nonumber\\
 2k_{(Y)} & = & -
        \frac{1}{3}\kappa^2a^2\tau_{00},
        \nonumber\\
 2k_{(L)0} & = & -\kappa^2\tau_{(L)0},
        \nonumber\\
 2k_{(LL)} & = & -\kappa^2\tau_{(LL)},
	\label{eqn:junction-cond}
  \end{eqnarray}
where $\tau_{00}$, $\tau_{(Y)}$, $\tau_{(L)0}$ and $\tau_{(LL)}$ are
$4$-gauge-invariant variables constructed in
ref.~\cite{Mukohyama2000b} from perturbations $\{t's\}$ of the surface
energy momentum tensor on the brane (see the last equation in
(\ref{eqn:harmonic-expansion})). The perturbed junction condition
(\ref{eqn:junction-cond}) was first derived in ref.~\cite{KIS} and 
rederived in ref.~\cite{Mukohyama2000c}.

The relation between the above $4$-gauge-invariant variables and 
gauge-invariant variables which are commonly used in the standard
cosmology~\cite{Bardeen,Kodama-Sasaki,MFB} is as follows. 
%
\begin{eqnarray}
 f_{00} & = & -2\Psi_{(KS)}, \nonumber\\
 f & = & 2a^2\Phi_{(KS)}, \nonumber\\
 \tau_{00} & = & 
        \rho\Delta_{(KS)} 
        + \dot{\rho}
        \frac{aV_{(KS)}}{|{\bf k}|}, \nonumber\\
 \tau_{(Y)} & = & a^2\left[p\Gamma_{(KS)}+c^2_s
	\left(\rho\Delta_{(KS)}  + \dot{\rho}
        \frac{aV_{(KS)}}{|{\bf k}|}\right)\right],
        \nonumber\\
 \tau_{(L)0} & = & (\rho+p)
        \frac{aV_{(KS)}}{|{\bf k}|}, \nonumber\\
 \tau_{(LL)} & = & a^2p \frac{\Pi_{(KS)}}{2{\bf k}^2}, 
	\label{eqn:relation}
 \end{eqnarray}
where variables with the subscript $(KS)$ are those defined in
ref.~\cite{Kodama-Sasaki} and $c_s$ is the sound velocity defined by
$c^2_s=\dot{p}/\dot{\rho}$.

\subsection{Master equation in the bulk}

As shown first in ref.\cite{Mukohyama2000b} and confirmed in 
ref.~\cite{KIS}\footnote{
In the latter paper, they extended the master equation of vector and 
tensor perturbations in the bulk to more general background without 
maximal symmetry.}, all components of perturbed Einstein's equation 
in the bulk around the pure AdS 
background can be reduced to a single equation called the master
equation. For $K=0$ and ${\bf k}\ne 0$, the master equation is of the
form  
%
\begin{equation}
 {\bf L}_{\bf k}\Phi = 0,
	\label{eqn:master-eq}
\end{equation}
where $\Phi$ is the master variable depending only on the coordinates
$\{x^a\}$ in the two-dimensional orbit space and the linear differential 
operator ${\bf L}_{\bf k}$ is defined by 
%
\begin{equation}
 {\bf L}_{\bf k}\Psi \equiv 
	r^2\nabla^a\left[r^{-1}\nabla_a(r^{-1}\Psi)\right]
	- {\bf k}^2r^{-2}\Psi. 
\end{equation}
For example, in the
conformally flat coordinate system considered in
section~\ref{sec:background} the master equation becomes 
%
\begin{equation}
 X^{-2}\partial_X\left[X\partial_X(X\Phi)\right]
	- \partial_T^2\Phi - {\bf k}^2\Phi = 0,
\end{equation}
and is easily solved. The general solution is
%
\begin{equation}
 \Phi =  \frac{1}{X}\left\{
	e^{-i|{\bf k}|T}(C\ln X +D)
	+ \int_0^{\infty} d\mu
	e^{-i\sqrt{{\bf k}^2+\mu^2}T}
	[A(\mu)J_0(\mu X)+B(\mu)Y_0(\mu X)]
	+ (c.c)\right\},
	\label{eqn:general-sol}
\end{equation}
where $(c.c)$ represents the complex conjugate.

Actually, $5$-gauge-invariant variables for scalar perturbations are
written in terms of the master variable as
%
\begin{eqnarray}
 F_{ab} & = & \frac{1}{r}\left(\nabla_a\nabla_b\Phi 
        -\frac{2}{3}\nabla^2\Phi\gamma_{ab}
        + \frac{1}{3l^2}\Phi\gamma_{ab}\right),
        \nonumber\\
 F & = & \frac{r}{3}\left(\nabla^2\Phi-\frac{2}{l^2}\Phi\right).
\end{eqnarray}

For ${\bf k}=0$, the corresponding perturbation has the plane
symmetry and, thus, the generalized Birkoff's theorem guarantees that
the perturbed bulk geometry is a S-AdS spacetime. Hence, the 
perturbation corresponding to ${\bf k}=0$ is actually perturbation of 
the parameter $\mu$ in (\ref{eqn:S-AdS}) around $\mu=0$ and can be 
understood as
dark radiation~\cite{Mukohyama2000a}. Since the effect of non-zero
$\mu$ has already been analyzed non-perturbatively in
refs.~\cite{CGS,FTW,BDEL,Mukohyama2000a,Kraus,Ida}, in this paper we
shall concentrate on perturbations with non-zero ${\bf k}$. This 
treatment is, of course, justified since perturbations with different
${\bf k}$ are decoupled from each other in the linear order.


\section{Equations on the brane}
	\label{sec:eq-on-brane}

As we have seen in the previous section, all $5$-gauge-invariant
variables and $4$-gauge-invariant variables are written in terms of
the master variable $\Phi$. After some cumbersome calculations using 
the master equation (\ref{eqn:master-eq}), we find the following set 
of three equations among the $4$-gauge-invariant geometrical
variables. 
%
\begin{eqnarray}
 \dot{k}_{(L)0} + 3Hk_{(L)0} - k_{00}
	+ 2a^{-2}k_{(Y)}
	+ \frac{l\dot{H}}{2\sqrt{1+l^2H^2}}f_{00}
	+ \frac{4}{3}{\bf k}^2a^{-2}k_{(LL)}
	& = & 0,\nonumber\\
 \dot{k}_{(Y)} -Hk_{(Y)} 
	+ \frac{l\dot{H}}{2\sqrt{1+l^2H^2}}(\dot{f}-2Hf)
	+ \frac{1}{3}{\bf k}^2k_{(L)0} + a^2Hk_{00}
	& = & 0,
	\label{eqn:conservation-eq_pre}
\end{eqnarray}
and
%
\begin{eqnarray}
 \ddot{f} + a^2H\dot{f}_{00} + 2(2H^2+\dot{H})(a^2f_{00}-f)
	+\frac{{\bf k}^2}{3a^2}(2f-a^2f_{00}) & &\nonumber\\
	-\frac{2a^2\sqrt{1+l^2H^2}}{l}k_{00}
	+\frac{2(3+3l^2H^2+l^2\dot{H})}{l\sqrt{1+l^2H^2}}k_{(Y)}
	& = & 0.
	\label{eqn:new-eq}
\end{eqnarray}
Note that the equations (\ref{eqn:conservation-eq_pre}) represent 
nothing but the gauge-invariant expression of perturbation of the
equation 
%
\begin{equation}
 \nabla^{(4)}_{\nu}(K^{\nu}_{\mu}-K\delta^{\nu}_{\mu})
	= R_{MN}e^M_{\mu}n^N = 0,
\end{equation}
where $\nabla^{(4)}_{\mu}$ is the four-dimensional covariant
derivative compatible with the induced metric $q_{\mu\nu}$. On the
other hand, (\ref{eqn:new-eq}) is the gauge-invariant expression of
perturbation of the equation~\footnote{The author thanks W. Israel for
pointing this out.}
%
\begin{equation}
 K^2-K^{\mu}_{\nu}K^{\nu}_{\mu}-R^{(4)}=2G_{MN}n^Mn^N=12l^{-2},
\end{equation}
and, as we shall see, can be interpreted as an evolution equation of
$f$ after use of the junction condition. Here, $R^{(4)}$ is the Ricci
scalar of the four-dimensional induced metric
$q_{\mu\nu}$. Furthermore, we can express the master variable
and its derivative at any point on the brane in terms of the
$4$-gauge-invariant geometrical variables. For example, 
%
\begin{eqnarray}
 \dot{f}-2Hf + \frac{{\bf k}^2}{3a^2H}f + a^2Hf_{00}
	+ \frac{2\sqrt{1+l^2H^2}}{lH}k_{(Y)}	
	& = & \frac{{\bf k}^4}{9a^3H}
	\left.\Phi\right|_{x^a=Z^{(0)a}(t)},\nonumber\\
	k_{(Y)}+a^2Hk_{(L)0}+\frac{2}{3}{\bf k}^2k_{(LL)}
	& = & \frac{{\bf k}^2}{6a}
	\left.r\partial_{\perp}(r^{-1}\Phi)\right|_{x^a=Z^{(0)a}(t)},
	\label{eqn:Phi-dnPhi}
\end{eqnarray}
where $\partial_{\perp}=n^a\partial_a$. If one like, one can
check these equations (\ref{eqn:conservation-eq_pre}),
(\ref{eqn:new-eq}) and (\ref{eqn:Phi-dnPhi}) by using the general 
solution (\ref{eqn:general-sol}) and a computer algebra package like
{\it GRTensor II}~\cite{GRTensorII}.

By substituting the doubly gauge-invariant junction condition
(\ref{eqn:junction-cond}), we obtain the perturbed conservation
equation 
%
\begin{eqnarray}
 \dot{\tau}_{(L)0} + 3H\tau_{(L)0} - \frac{1}{a^2}\tau_{(Y)}
	+ \frac{4}{3}\frac{{\bf k}^2}{a^2}\tau_{(LL)}
	- \frac{l\dot{H}}{\kappa_5^2\sqrt{1+l^2H^2}}f_{00}
	& = & 0,\nonumber\\
 \dot{\tau}_{00} +3H\tau_{00}+ \frac{3H}{a^2}\tau_{(Y)}
	+ \frac{{\bf k}^2}{a^2}\tau_{(L)0} 
	- \frac{3l\dot{H}}{\kappa_5^2a^2\sqrt{1+l^2H^2}}
	(\dot{f}-2Hf)
	& = & 0,
	\label{eqn:conservation-eq}
\end{eqnarray}
the evolution equation of $f$
%
\begin{eqnarray}
 \ddot{f} + a^2H\dot{f}_{00} & + & 2(2H^2+\dot{H})(a^2f_{00}-f)
	+\frac{{\bf k}^2}{3a^2}(2f-a^2f_{00}) \nonumber\\
	& + & \frac{\kappa_5^2\sqrt{1+l^2H^2}}{l}\tau_{(Y)}
	-\frac{\kappa_5^2a^2(1+l^2H^2+l^2\dot{H})}
	{3l\sqrt{1+l^2H^2}}\tau_{00}
	= 0,
	\label{eqn:eq-of-f}
\end{eqnarray}
and the expressions of $\Phi$ and $r\partial_{\perp}(r^{-1}\Phi)$ on
the brane 
%
\begin{eqnarray}
 \dot{f}-2Hf + \frac{{\bf k}^2}{3a^2H}f + a^2Hf_{00}
	- \frac{1}{3}\kappa_5^2a^2\frac{\sqrt{1+l^2H^2}}{lH}
	\tau_{00}
	& = & \frac{{\bf k}^4}{9a^3H}
	\left.\Phi\right|_{x^a=Z^{(0)a}(t)},
	\label{eqn:Phi-on-brane}\\
 \tau_{00}+3H\tau_{(L)0} +\frac{2{\bf k}^2}{a^2}\tau_{(LL)}
	& = & -\frac{{\bf k}^2}{\kappa_5^2a^3}
	\left.r\partial_{\perp}(r^{-1}\Phi)\right|_{x^a=Z^{(0)a}(t)}.
	\label{eqn:dnPhi-on-brane}
\end{eqnarray}

Now we shall use the equation (\ref{eqn:dnPhi-on-brane}) to give a
boundary condition to $\Phi$ and use another equation
(\ref{eqn:Phi-on-brane}) to give a feedback of $\Phi$ to perturbations 
on the brane. First, let us introduce the $Z_2$-symmetric retarded
Green's function $G_{\bf R}(x^a,{x'}^a)$ satisfying
%
\begin{equation}
 {\bf L}_{\bf k}G_{\bf R}(x^a,{x'}^a) = 
	\frac{\delta^2(x^a-{x'}^a)}{\sqrt{-\gamma}}
\end{equation}
in the ${\bf Z}_2$-symmetrized orbit space. Here, the 
${\bf Z}_2$-symmetrized orbit space is obtained from two copies of 
the region (i) of the AdS spacetime glued together along the 
unperturbed world volume $\Sigma^{(0)}$ of the brane (see 
section~\ref{sec:background}). Hence, $\Phi$ is expressed as 
%
\begin{equation}
 \Phi(x^a) = \Phi_{in}(x^a)-2\int dt'\ 
	G_{\bf R}(x^a,Z^{(0)a}(t'))
	\ \left[\frac{\kappa_5^2a(t')^3}{{\bf k}^2}S(t')
	+\left.r\partial_{\perp}(r^{-1}\Phi_{in})
	\right|_{x^a=Z^{(0)a}(t')}\right],
\end{equation}
where $\Phi_{in}(x^a)$ is an arbitrary solution of the master 
equation (\ref{eqn:master-eq}) representing initial gravitational 
waves in the bulk and $S(t)$ is the left hand side of 
(\ref{eqn:dnPhi-on-brane}). Hence, by substituting this into 
(\ref{eqn:Phi-on-brane}) we obtain
%
\begin{eqnarray}
 & R(t)& - \frac{{\bf k}^4}{9a(t)^3H(t)}\Phi_{in}(Z^{(0)a}(t)) 
	\nonumber\\
 & + &	\frac{2{\bf k}^2}{9a(t)^3H(t)} \int dt'\ 
	\bar{G}_{\bf R}(t,t')
	\ \left[\kappa_5^2a(t')^3S(t')
	+{\bf k}^2\left.r\partial_{\perp}(r^{-1}\Phi_{in})
	\right|_{x^a=Z^{(0)a}(t')}\right] = 0,
	\label{eqn:integro-diff-eq}
\end{eqnarray}
where $\bar{G}_{\bf R}(t,t')=G_{\bf R}(Z^{(0)a}(t),Z^{(0)a}(t'))$ and
$R(t)$ is the left hand side of (\ref{eqn:Phi-on-brane}). 
This is the integro-differential equation which we have been
seeking since conjectured in ref.~\cite{Mukohyama2000b}. If
there is no initial gravitational waves in the bulk, then the
integro-differential equation is reduced to 
%
\begin{equation}
 R(t) + \frac{2\kappa_5^2{\bf k}^2}{9a(t)^3H(t)} \int dt'\ 
	\bar{G}_{\bf R}(t,t')
	\ a(t')^3S(t') = 0.
\end{equation}

Finally, we have found the set of four coupled equations
(\ref{eqn:conservation-eq}), (\ref{eqn:eq-of-f}) and
(\ref{eqn:integro-diff-eq}). These equations are written in terms of
physical quantities on the brane and can be used to determine the
evolution of cosmological perturbations on the brane. Note that the 
number of equations, four, is the same as the number of independent
components of perturbed Einstein's equation for scalar perturbations
in the standard cosmology as we shall see in the next section.

Although we have succeeded in deriving the integro-differential
equation formally, we need an explicit expression of the ${\bf
Z}_2$-symmetric retarded Green's function in order to obtain the
kernel. Since evaluation of the Green's function for a general motion
of the brane seems difficult, it might be more realistic to adopt
numerical methods to integrate the master equation 
(\ref{eqn:master-eq}) in the bulk and the three differential equations
on the brane, (\ref{eqn:conservation-eq}) and (\ref{eqn:eq-of-f}), at
the same time. In this case, we can use the equations 
(\ref{eqn:dnPhi-on-brane}) and (\ref{eqn:Phi-on-brane}) to give a
boundary condition to $\Phi$ and to give a feedback of $\Phi$ to
perturbations on the brane.


\section{Comparison with the standard cosmology}
	\label{sec:comparison}

In this section we compare the four independent equations derived in 
the previous section with the corresponding equations in the 
standard cosmology.

\subsection{Brane-world cosmology}

By using the background junction condition
(\ref{eqn:Friedmann-eq-pre}) with the plus sign and the conservation
equation (\ref{eqn:background-conservation}), we can rewrite the
four equations obtained in the previous section as follows. The
conservation equation is 
%
\begin{eqnarray}
 \dot{\tau}_{(L)0} + 3H\tau_{(L)0} - \frac{1}{a^2}\tau_{(Y)}
	+ \frac{4}{3}\frac{{\bf k}^2}{a^2}\tau_{(LL)}
	+ \frac{1}{2}(\rho+p)f_{00}
	& = & 0,\nonumber\\
 \dot{\tau}_{00} +3H\tau_{00}+ \frac{3H}{a^2}\tau_{(Y)}
	+ \frac{{\bf k}^2}{a^2}\tau_{(L)0}
	+ \frac{3}{2a^2}(\rho+p)(\dot{f}-2Hf)
	& = & 0,
	\label{eqn:brane-conservation-eq}
\end{eqnarray}
the evolution equation of $f$ is
%
\begin{eqnarray}
 \ddot{f} + a^2H\dot{f}_{00} & + & 2(2H^2+\dot{H})(a^2f_{00}-f)
	+\frac{{\bf k}^2}{3a^2}(2f-a^2f_{00}) \nonumber\\
	& + & 8\pi G_N\left(1+\frac{\rho}{\lambda}\right)
	\tau_{(Y)} 
	-\frac{8\pi G_N}{3}\left(1-\frac{2\rho+3p}{\lambda}\right)
	a^2\tau_{00}
	= 0,
	\label{eqn:brane-f-eq}
\end{eqnarray}
and the integro-differential equation without initial gravitational
waves in the bulk, $\Phi_{in}=0$, is
%
\begin{equation}
 R(t) + \frac{2\kappa_5^2{\bf k}^2}{9a(t)^3H(t)} \int dt'\ 
	\bar{G}_{\bf R}(t,t')
	\ a(t')^3S(t') = 0.
	\label{eqn:brane-integro-diff}
\end{equation}
where 
%
\begin{eqnarray}
 R(t) & = & 
	\dot{f}-2Hf + \frac{{\bf k}^2}{3a^2H}f + a^2Hf_{00}
	- \frac{8\pi G_N}{3}\left(1+\frac{\rho}{\lambda}\right)
	\frac{a^2}{H}\tau_{00},
	\label{eqn:R(t)} \\
 S(t) & = & 
	\tau_{00}+3H\tau_{(L)0} +\frac{2{\bf k}^2}{a^2}\tau_{(LL)}. 
	\label{eqn:S(t)}
\end{eqnarray}
Here, we have used the definition of four-dimensional effective
Newton's constant $G_N$ and the four-dimensional effective
cosmological constant $\Lambda_4$ introduced in
section~\ref{sec:background}.

\subsection{Standard cosmology}

Now, for the purpose of the comparison, let us review the
gauge-invariant formalism of cosmological perturbations in the
four-dimensional Einstein gravity~\cite{Bardeen,Kodama-Sasaki,MFB} 
in terms of the $4$-gauge-invariant variables used in the present
paper. We consider scalar perturbations around the 
Friedmann-Robertson-Walker spacetime. Namely, we consider the 
metric
%
\begin{equation}
 ds_4^2 = (q^{(0)}_{\mu\nu}+\delta q_{\mu\nu})dy^{\mu}dy^{\nu},
\end{equation}
where 
%
\begin{equation}
 q^{(0)}_{\mu\nu}dy^{\mu}dy^{\nu} 
	= -dt^2 + a^2(t)\sum_{i=1}^3(dy^i)^2,
\end{equation}
and 
%
\begin{equation}
 \delta q_{\mu\nu}dy^{\mu}dy^{\nu} 
	= \int d^3{\bf k}\left[
        \sigma_{00}Ydt^2
        + 2\sigma_{(L)0}V_{(L)i}dtdy^i
        + (\sigma_{(LL)}T_{(LL)ij}
	+\sigma_{(Y)}T_{(Y)ij})dy^idy^j\right]. 
\end{equation}
Here, the coefficients $\sigma_{00}$, $\sigma_{(L)0}$,
$\sigma_{(LL)}$, $\sigma_{(Y)}$ depend only on the proper time
$t$. Following ref.~\cite{Mukohyama2000c}, we can construct
gauge-invariant variables $f_{00}$ and $f$. 
%
\begin{eqnarray}
 f_{00} & = & \sigma_{00} 
        - 2\dot{\chi},  \nonumber\\
 f & = & \sigma_{(Y)} + 2a^2H\chi
        + \frac{2{\bf k}^2}{3}\sigma_{(LL)},
\end{eqnarray}
where $\chi=
\sigma_{(L)0}-\dot{\sigma}_{(LL)}+2H\sigma_{(LL)}$.

As for the stress energy tensor, we consider
%
\begin{equation}
 T^{\mu}_{(4)\nu} = -\frac{\Lambda_4}{8\pi G_N}\delta^{\mu}_{\nu} +
	T^{\mu}_{\nu} + \delta \tilde{T}^{\mu}_{\nu},
\end{equation}
where $\Lambda_4$ is the four-dimensional cosmological constant,
$T_{\mu\nu}$ is given by (\ref{eqn:Tmunu}) and 
%
\begin{equation}
 \delta \tilde{T}_{\mu\nu}dy^{\mu}dy^{\nu} 
	= \int d^3{\bf k}\left[
        \tilde{t}_{00}Ydt^2
        + 2\tilde{t}_{(L)0}V_{(L)i}dtdy^i
        + (\tilde{t}_{(LL)}T_{(LL)ij}
	+\tilde{t}_{(Y)}T_{(Y)ij})dy^idy^j\right]. 
\end{equation}
Again, the coefficients $\tilde{t}_{00}$, $\tilde{t}_{(L)0}$,
$\tilde{t}_{(LL)}$, $\tilde{t}_{(Y)}$ depend only on the proper 
time $t$. Following ref.~\cite{Mukohyama2000b}, we can construct
gauge-invariant variables $\tau_{00}$, $\tau_{L(0)}$, $\tau_{(LL)}$
and $\tau_{(Y)}$. 
%
\begin{eqnarray}
 \tau_{00} & = & \tilde{t}_{00} + \dot{\rho}\chi, 
        \nonumber\\
 \tau_{(L)0} & = & \tilde{t}_{(L)0}
        + \frac{1}{2}(\rho+p)
        \left[\chi-\dot{\sigma}_{(LL)}+2H\sigma_{(LL)}\right],
        \nonumber\\
 \tau_{(LL)} & = & \tilde{t}_{(LL)}
        \nonumber\\
 \tau_{(Y)} & = & \tilde{t}_{(Y)} + a^2\dot{p}\chi. 
\end{eqnarray}

We can expand the four-dimensional Einstein's equation
$G^{\mu}_{(4)\nu}=8\pi G_NT^{\mu}_{(4)\nu}$ up to the first order in
the perturbations, where $G^{\mu}_{(4)\nu}$ is the Einstein tensor
constructed from the metric $q_{\mu\nu}$, and express it in terms of
the above gauge-invariant variables. The background equations are, of
course, the Friedmann equation 
%
\begin{equation}
 H^2 = \frac{8\pi G_N}{3}\rho + \frac{\Lambda_4}{3},
\end{equation}
and the conservation equation
%
\begin{equation}
 \dot{\rho} + 3H(\rho+p) = 0. 
\end{equation}
The first order equations are 
%
\begin{eqnarray}
 \dot{f} -2Hf +\frac{{\bf k}^2}{3a^2H}f
	+ a^2Hf_{00} & = & \frac{8\pi G_N}{3H}a^2\tau_{00},
	\nonumber\\
 \dot{f} -2Hf + a^2Hf_{00} & = & -8\pi G_Na^2\tau_{(L)0}, \nonumber\\
 a^2f_{00}-f & = & 32\pi G_Na^2\tau_{(LL)},
	\nonumber\\
 \ddot{f} - H\dot{f} - 2(H^2+\dot{H})f
	+ \frac{{\bf k}^2}{3a^2}f
	+ a^2H\dot{f}_{00} + a^2(3H^2+2\dot{H})f_{00}
	- \frac{{\bf k}^2}{3}f_{00}
	& = & -8\pi G_N\tau_{(Y)}.
\end{eqnarray}
These four equations are equivalent to the following set of four
equations. 
%
\begin{eqnarray}
 \dot{\tau}_{(L)0} + 3H\tau_{(L)0} - \frac{1}{a^2}\tau_{(Y)}
	+ \frac{4}{3}\frac{{\bf k}^2}{a^2}\tau_{(LL)}
	+ \frac{1}{2}(\rho+p)f_{00}
	& = & 0,\nonumber\\
 \dot{\tau}_{00} +3H\tau_{00}+ \frac{3H}{a^2}\tau_{(Y)}
	+ \frac{{\bf k}^2}{a^2}\tau_{(L)0}
	+ \frac{3}{2a^2}(\rho+p)(\dot{f}-2Hf)
	& = & 0, 
	\label{eqn:standard-conserved-eq}
\end{eqnarray}
%
\begin{eqnarray}
 \ddot{f} + a^2H\dot{f}_{00} & + & 2(2H^2+\dot{H})(a^2f_{00}-f)
	+\frac{{\bf k}^2}{3a^2}(2f-a^2f_{00}) \nonumber\\
 & + & 8\pi G_N\tau_{(Y)} - \frac{8\pi G_N}{3}a^2\tau_{00} = 0,
	\label{eqn:standard-f-eq}
\end{eqnarray}
and
%
\begin{equation}
 \dot{f} -2Hf +\frac{{\bf k}^2}{3a^2H}f
	+ a^2Hf_{00} - \frac{8\pi G_N}{3}\frac{a^2}{H}
	\tau_{00} = 0. 
	\label{eqn:standard-integro-diff}
\end{equation}

Finally, in the standard cosmology, there are two well-known algebraic 
relations among gauge-invariant variables derived from the perturbed 
Einstein's equation. With our notation, those are 
%
\begin{eqnarray}
 {\bf k}^2f & = & 8\pi G_Na^4(\tau_{00}+3H\tau_{(L)0}),
	\nonumber\\
  a^2f_{00}-f & = & 32\pi G_Na^2\tau_{(LL)}. 
	\label{eqn:standard-algebraic}
\end{eqnarray}

\subsection{Similarity and difference}

The equations (\ref{eqn:standard-conserved-eq}) in the standard
cosmology are nothing but the perturbed conservation equation and, of
course, exactly the same as those for the brane-world cosmological
perturbations (\ref{eqn:brane-conservation-eq}). The equation 
(\ref{eqn:standard-f-eq}) in the standard cosmology is similar to the
equation (\ref{eqn:brane-f-eq}) in the brane-world cosmology. 
In particular, in the low energy limit when $|\rho|/\lambda\ll 1$ and
$|p|/\lambda\ll 1$, these equations become identical. The final
equation (\ref{eqn:standard-integro-diff}) in the standard cosmology
is, in some sense, similar to the integro-differential equation
(\ref{eqn:brane-integro-diff}). Actually, in a situation when the
effect of the bulk gravitational waves (the integral in the
integro-differential equation) is negligible~\footnote{Actually, in
the static background $H=0$, the so called zero-mode
truncation~\cite{Garriga-Tanaka} corresponds to setting 
$\bar{G}_{\bf R}(t,t')\simeq 0$}, the integro-differential equation is
reduced to $R(t)\simeq 0$, which becomes the same as
(\ref{eqn:standard-integro-diff}) in the low energy limit.

However, in general, non-local properties of the integro-differential 
equation (\ref{eqn:brane-integro-diff}) in the brane-world cosmology
are expected to become significant. Moreover, in high energy,
corrections of order $O(\rho/\lambda)$ and $O(p/\lambda)$ become
significant, too. In particular, in the brane-world cosmology we have
not found simple algebraic relations like
(\ref{eqn:standard-algebraic}). Hence, there is a good possibility
that we can find observable differences between the brane-world
cosmology and the standard cosmology, by analyzing
the set of equations (\ref{eqn:brane-integro-diff}),
(\ref{eqn:brane-f-eq}) and
(\ref{eqn:brane-conservation-eq}). Furthermore, if we consider the 
initial gravitational waves in the bulk ($\Phi_{in}\ne 0$) then these
waves affect the evolution of perturbations on the brane as expressed
by the more general integro-differential equation
(\ref{eqn:integro-diff-eq}).


\section{Summary}
	\label{sec:summary}

We have analyzed cosmological perturbations in the Randall-Sundrum
brane-world scenario with a positive tension brane in the AdS
background bulk geometry by using the doubly gauge-invariant
formalism. For simplicity, we have considered scalar perturbations
around a plane symmetric ($K=0$) background. We have derived the set 
of four coupled equations (\ref{eqn:conservation-eq}),
(\ref{eqn:eq-of-f}) and (\ref{eqn:integro-diff-eq}). The final
equation is an integro-differential equation whose kernel is
constructed formally from the ${\bf Z}_2$-symmetric retarded Green's
function of the bulk gravitational waves. Hence, this equation
describes non-local properties due to the gravitational waves
propagating in the bulk. Finally, we compare these equations with
the corresponding equations in the standard cosmology. In particular,
we have seen that the number of independent equations, four, is the
same as in the standard cosmology and that in low energy these sets of
equations differ only by the non-local effects due to gravitational
waves in the bulk.

Although we have succeeded in deriving the integro-differential
equation formally, we need an explicit expression of the ${\bf
Z}_2$-symmetric retarded Green's function in order to obtain the
kernel. Since evaluation of the Green's function for a general motion
of the brane seems difficult, it might be more realistic to adopt
numerical methods to integrate the master equation
(\ref{eqn:master-eq}) in the bulk and the three differential equations
on the brane, (\ref{eqn:conservation-eq}) and (\ref{eqn:eq-of-f}), at
the same time. In this case, we can use the equations 
(\ref{eqn:dnPhi-on-brane}) and (\ref{eqn:Phi-on-brane}) to give a
boundary condition to $\Phi$ and to give a feedback of $\Phi$ to
perturbations on the brane. Namely, we have one equation for bulk
perturbations, three equations for perturbations on the brane and two
equations to relate these two kinds of perturbations.

\vspace{1cm}

The author would like to thank Werner Israel for continuing
encouragement and helpful discussions. This work was supported by the
CITA National Fellowship and the NSERC operating research grant.



\begin{references}
\bibitem{RS2}
L.~Randall and R.~Sundrum, Phys. Rev. Lett. {\bf 83}, 4690 (1999)
[hep-th/9906064]. 
\bibitem{CGS}
J.~M.~Cline, C.~Grojean and G.~Servant, Phys. Rev. Lett. 
{\bf 83} 4245 (1999) [hep-ph/9906523].
\bibitem{FTW}
E.~E.~Flanagan, S.~H.~H.~Tye, I.~Wasserman, Phys. Rev. {\bf D62},
044039 (2000) [hep-ph/9910498].
\bibitem{BDEL}
P.~Bin\'{e}truy, C.~Deffayet, U.~Ellwanger and D.~Langlois,
Phys. Lett. {\bf B477}, 285 (2000) [hep-th/9910219]. 
\bibitem{Mukohyama2000a}
S.~Mukohyama, Phys. Lett. {\bf B473}, 241 (2000) [hep-th/9911165]. 
\bibitem{Kraus}
P.~Kraus, JHEP {\bf 9912}, 011 (1999) [hep-th/9910149].
\bibitem{Ida}
D.~Ida, JHEP {\bf 0009}, 014 (2000) [gr-qc/9912002].
\bibitem{Mukohyama2000b}
S.~Mukohyama, Phys. Rev. {\bf D62}, 084015 (2000) [hep-th/0004067].
\bibitem{Mukohyama2000c}
S.~Mukohyama, Class. Quant. Grav. {\bf 17}, 4777 (2000)
[hep-th/0006146]. 
\bibitem{KIS}
H.~Kodama, A.~Ishibashi and O.~Seto, Phys. Rev. {\bf D62}, 064022
(2000) [hep-th/0004160].
\bibitem{Kodama}
H.~Kodama, hep-th/0012132. 
\bibitem{Maartens}
R.~Maartens, Phys. Rev. {\bf D62}, 084023 (2000) [hep-th/0004166]; 
C.~Gordon and R.~Maartens, Phys. Rev. {\bf D63}, 044022 (2001)
[hep-th/0009010]; 
R.~Maartens, V.~Sahni and T.~D.~Saini, Phys. Rev. {\bf D63}, 063509
(2001) [gr-qc/0011105].
\bibitem{Langlois}
D.~Langlois, Phys. Rev. {\bf D62}, 126012 (2000) [ hep-th/0005025]; 
Phys. Rev. Lett. {\bf 86},  2212 (2001) [hep-th/0010063].
\bibitem{BDBL}
C.~van~de~Bruck, M.~Dorca, R.~Brandenberger and A.~Lukas,
Phys. Rev. {\bf D62}, 123515 (2000) [hep-th/0005032];
C. van~de~ Bruck and M.~Dorca, hep-th/0012073; 
M.~Dorca and C.~van~de~Bruck, hep-th/0012116.
\bibitem{Koyama-Soda}
K.~Koyama and J.~Soda, Phys. Rev. {\bf D62}, 123502 (2000) [hep]
\bibitem{LMW}
D.~Langlois, R.~Maartens and D.~Wands, Phys. Lett. {\bf B489}, 259
(2000) [hep-th/0006007].
\bibitem{LMSW}
D.~Langlois, R.~Maartens, M.~Sasaki and David Wands, Phys. Rev. {\bf
D63}, 084009 (2001) [hep-th/0012044].
\bibitem{MSM}
S.~Mukohyama, T.~Shiromizu and K.~Maeda, Phys. Rev. {\bf D62}, 024028
(2000), Erratum-ibid. {\bf D63}, 029901 (2001) [hep-th/9912287].
\bibitem{Birmingham}
D.~Birmingham, Class. Quant. Grav. {\bf 16}, 1197(1999)
[hep-th/9808032]. 
\bibitem{Hawking}
S.~W.~Hawking, Commun. Math. Phys. {\bf 43}, 199 (1975).
\bibitem{GW}
W.~D.~Goldberger and M.~B.~Wise, Phys. Rev. Lett. {\bf 83}, 4922
(1999) [hep-ph/9907447]. 
\bibitem{DFGK}
O.~DeWolfe, D.~Z.~Freedman, S.~S.~Gubser and A.~Karch, Phys. Rev. {\bf
D62}, 046008 (2000) [hep-th/9909134]. 
\bibitem{Mukohyama2001a}
S.~Mukohyama, Phys. Rev. {\bf D63}, 044008 (2001) [hep-th/0007239]. 
\bibitem{Mukohyama2001b}
S.~Mukohyama, Phys. Rev. {\bf D}, to appear [hep-th/0101038]. 
\bibitem{Israel}
W.~Israel, Nuovo Cim. {\bf B44}, 1 (1966); Erratum-ibid. {\bf B48},
463 (1967). 
\bibitem{SMS}
T.~Shiromizu, K.~Maeda and M.~Sasaki, Phys. Rev. {\bf 62}, 024012
(2000) [gr-qc/9910076].
\bibitem{TM}
T.~Tanaka and X.~Montes, Nucl. Phys. {\bf B582}, 259 (2000)
[hep-th/0001092]. 
\bibitem{Bardeen}
J.~M.~Bardeen, Phys. Rev. {\bf D22}, 1882 (1980). 
\bibitem{Kodama-Sasaki}
H.~Kodama and M.~Sasaki, Prog. Theor. Phys. Suppl. {\bf 78}, 1
(1984). 
\bibitem{MFB}
V.~F.~Mukhanov, H.~A.~Feldman, and R.~H.~Brandenberger, Phys. Rep. 
{\bf 215}, 203 (1992). 
\bibitem{GRTensorII}
{\it GRTensor II}, http://grtensor.phy.queensu.ca .
\bibitem{Garriga-Tanaka}
J.~Garriga and T.~Tanaka, Phys. Rev. Lett. {\bf 84}, 2778 (2000)
[hep-th/9911055]. 
\end{references}
\end{document}